# Pulse Compression by an Optical Push Broom On a Chip

Boyi Zhang[1], Maurice Pfeiffer[1], Mahmoud A. Gaafar[1,2,3], He Li[4], Xinlun Cai[4], Juntao Li[4*], Manfred Eich[1,5], and Alexander Yu. Petrov[1,5†]

[1]Institute of Optical and Electronic Materials, Hamburg University of Technology, Hamburg 21073, Germany
[2]Department of Physics, Faculty of Science, Menoufia University, Menoufia, Egypt
[3]Directed Energy Research Centre, Technology Innovation Institute, Abu Dhabi, SE45-01, United Arab Emirates
[4]State Key Laboratory of Optoelectronic Materials & Technology, Sun Yat-sen University, Guangzhou 510275, China
[5]Institute of Functional Materials for Sustainability, Helmholtz-Zentrum Hereon, Max-Planck-Strasse 1, Geesthacht, D-21502, Germany

*lijt3@mail.sysu.edu.cn
†a.petrov@tuhh.de

**Abstract**

In this study, we report a first experimental demonstration of optical pulse compression by a gradual refractive index front co-moving in a periodically modulated silicon waveguide, the so-called optical push broom effect. Optical push broom captures and confines the input signal pulse inside a faster propagating refractive index front, driven by a pump pulse. This is a spatio-temporal analogue of light trapping in a tapered plasmonic waveguide where light is continuously changing its wavevector approaching zero group velocity and, thus, stopped without reflection. Here the signal is captured and accelerated by the front until its velocity matches that of the front, effectively stopping the light relative to the front. To model this phenomenon, we employ the slowly varying envelope approximation. Notably, we successfully reproduced the experimental frequency shift at the output corresponding to the temporal delay at the input.

## 1. Introduction

Interest in light interactions with time-varying media is growing due to their capability for manipulation of not only the wavenumber but also the frequency of light [1–3]. In waveguide and photonic crystal systems additional flexibility is added by the possibility to design dispersion [4–

8]. Specifically, moving fronts can capture and compress light pulses, offering a promising approach for one-step pulse compression [9–13]. The interaction with the index front can be considered as an indirect transition of the signal state, altering its frequency and wavenumber [14,15]. Depending on dispersion of the band diagram and the direction of the indirect transition, the signal is shifted to the perturbed states, behind the front, or back to the original states, before the front, which results in the transmission or reflection from the front [16–20]. However, if the transition goes parallel in between perturbed and original dispersion relations, the trapping of the pulse is realized as signal does not find states before and after the front and is bound to stay inside it [20]. In other words, the front captures and accelerates the signal pulse up to the front speed, whereas frequency and wavenumber of the signal are continuously changing as compression progresses with increasing propagation time and distance of signal and pump travelling together. This effect is named optical push broom effect [10]. The push broom effect shares conceptual similarities with the light localization by tapered plasmonic waveguides, where the geometry-driven index fronts works as the purely spatio analogue of moving fronts [21].

De Sterke first predicted that the energy of a weak probe could be made to heap up on the pump's leading edge propagating in a fiber Bragg grating [22]. Later, the Schrödinger equation description was given to describe the optical push broom effect involving cross phase modulation in highly dispersive media [23]. Recently Gaafar *et al*. have demonstrated the analytical solutions for pulse compression in a linear front [24], where a perturbed dispersion relation is piecewise linear and is shifted in frequency in respect to original dispersion relation. In this case, finite compression of the signal is achieved, which resembles focusing by a lens. On the other hand, Pendry has presented an analytical model to describe the compression of signal pulses interacting with a refractive index front that leads to the change of the slope of dispersion relation [13]. In this case, compression is continuous and might lead to singularities. Yet, experimental progress has lagged behind theoretical developments due to the complexity of the phenomena. Broderick *et al.* has reported the attempt to realize the optical push broom in fiber Bragg gratings by Kerr effect [10]. They showed a small portion of a CW signal could be cut out and concentrated inside the index front. We have shown a similar effect in periodically modulated silicon waveguides using free carrier excitation as the front [25]. But we could not measure duration of the trapped portion of the CW light and thus could not confirm the compression. Different systems are utilized now to generate fronts in epsilon near zero media, but propagation lengths stay very short due to intrinsic absorption of such media [26,27].

In this study, we achieved the first experimental confirmation of the optical push broom effect in a silicon Bragg waveguide, demonstrated through a pump probe experiment by varying the delay between pump and probe. We took advantage of the silicon waveguide's nonlinearity [28] and customizable dispersion relation [29], compared to fiber-based Bragg gratings [10]. The induced refractive index change is much smaller than in epsilon near zero media [30], but much larger propagation distances are possible. A slowly varying envelope approximation is employed to model the underlying physical process. The front is generated via two photon absorption of a 2 ps pump pulse centered at a wavelength beyond the band gap of the Bragg waveguide. The 10 ps signal pulse is tuned close to the band edge and propagates initially slower than the pump. Our experimental results show that the signal pulse is collected by the front and converted into new frequencies, with an estimated temporal compression of up to 10 times. One step further, the delay-dependent measurements untangle the Fourier transformation performed by the front [12]. We show that the temporal delay at the input results into the frequency shift at the output, which is predicted by the analytical model and well reproduced by numerical simulations.

**2. Experiment**

## 2.1. Design and manufacture of silicon Bragg waveguide

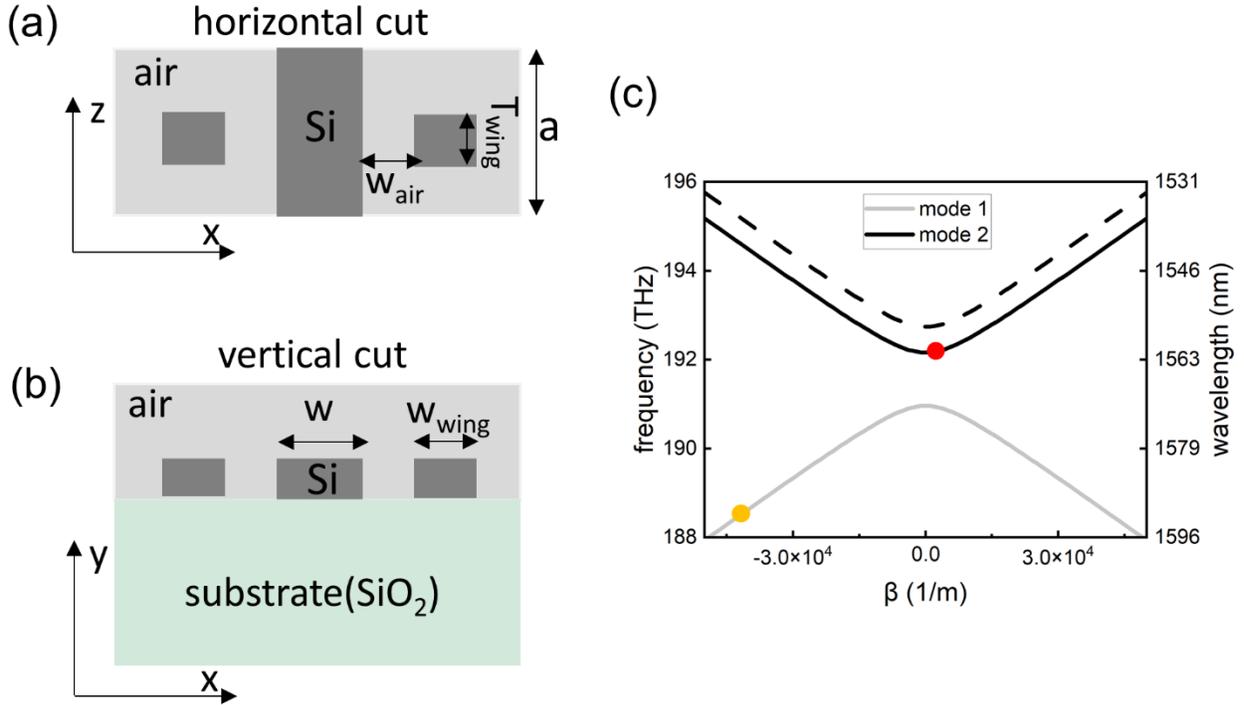

**Figure 1**: Schematic of the 1 mm-long silicon waveguide with a periodic perturbation: horizontal cut (a) and vertical cut (b). The silicon slab height and width are 220 nm and 500 nm, respectively. The height, width ($W_{wing}$), and the thickness $T_{wing}$ of the silicon wings are 220 nm, 220 nm and 150 nm, respectively. The air gap between the silicon waveguide and the wings is $W_{air} = 40\ nm$, and the lattice constant $a$ is 330 nm. (c) Band diagram of the periodic waveguide approximated by two hyperbolas. The black and grey curves correspond to the upper and lower bands and dashed curve to the perturbed upper band, whereas the perturbed lower band is not shown. Orange and red circles indicate the proposed pump and signal frequencies at the input, respectively.

In order to demonstrate the optical push broom effect on a chip, a silicon Bragg waveguide with periodic wings is designed. A schematic of the waveguide design is shown in Figure 1 (a-b). Figure 1(c) presents the band diagram of the waveguide which can be approximated by a hyperbolic curve which will be introduced in later numerical simulations. Solid black and grey lines represent the upper and lower branches of the dispersion relation, respectively, while the dashed black line represents the shifted band due to the presence of the refractive index front. Technically, had we positioned both the pump pulse and the signal wave on the upper branch of the dispersion relation, it would have been challenging to detect the blue shifted signal due to frequency overlap with the pump. Therefore, we positioned the pump frequency on the nondispersive part of the lower branch of the dispersion relation (orange circle in Figure 1(c)).

The employed silicon waveguide was fabricated on a silicon-on-insulator substrate with slab height of 220 nm and width of 500 nm. The patterns were exposed on a spin-coated AR-P 6200.13 electron-beam photoresist by using electron-beam lithography. Afterwards, it was transferred into the silicon slab by using an inductively coupled plasma dry etching. The residual photoresist was removed by the oxygen plasma etching and a microresist remover 1165. A SEM image of the fabricated waveguide is shown in Figure 2(a).

Light is coupled into and out of the Bragg grating using two grating couplers in combination with a single-mode fiber. Each grating coupler measures 20 µm in length and 12 µm in width and features a 400 µm adiabatic taper, gradually transitioning to a 500 nm-wide waveguide [31]. The structure of the grating is composed of air hole rows embedded in silicon. This coupler exhibits a 3 dB bandwidth of 30 nm, with an optimized peak coupling wavelength at 1570 nm and a minimum coupling loss of approximately 7 dB per coupler. Further details on the waveguide design and fabrication process can be found in the supporting information. The measured linear transmission (inclusive of coupling losses) and the group index of the TE-mode for the 1 mm-long silicon Bragg waveguide are presented in Figure 2(b) and 2(c). The overall shape of the transmission curve outside of the band gap region is due to the transmission characteristics of the grating coupler. The spikes in the transmission curve are caused by the Fabry-Perot oscillation at the input and output of the periodic waveguide. For group index measurements, the delay of a weak pulse with about

0.4 nm bandwidth is measured by an optical sampling oscilloscope. This measurement disregards the Fabry-Perot oscillations and tracks only the first transmitted pulse. As can be seen, the group index increases as the wavelength approaches the band edge.

**Figure 2**: Characteristics of the fabricated silicon Bragg waveguide. (a) SEM of the waveguide. Structure on the waveguide surface might come from the residual photoresist. It has weak effect on transmission due to small amplitude of the TE-mode fields at upper interface of the waveguide. (b) Normalized transmission in

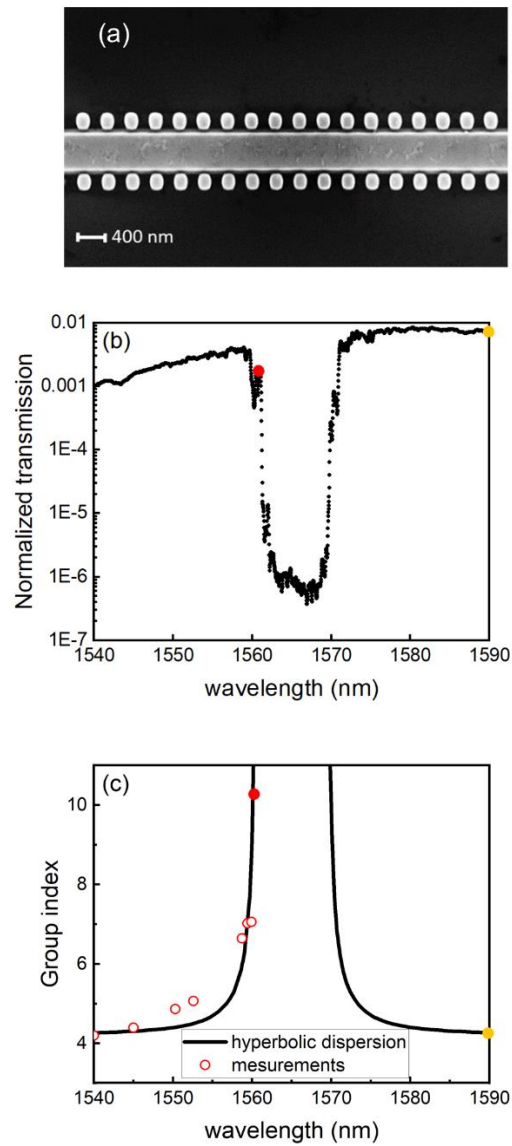

log scale. (c) Group index evaluated from pulse delay measurements (dots) and calculated from the hyperbolic dispersion curve approximation (black curve). Particularly, solid red and orange dots indicate the wavelengths chosen for experiments of the initial signal (1561 nm) and pump (1590 nm) wavelengths, respectively.

We have to mention also, that self-phase-modulation causes a temporally varying instantaneous frequency of the pump. This way, an initial non-chirped pump pulse acquires a frequency chirp. Still this frequency chirp does not significantly change the pulse envelope, thus, also the free carrier front shape as the pump is propagating at the frequency with negligible dispersion.

## 2.2. Pump and probe measurements

Optical push broom is experimentally demonstrated in the above chip with pump probe measurements. A mode-locked fiber laser (Menlo system) with a 100 MHz repetition rate delivers pulses of approx. 100 fs duration with a wide spectrum ranging from 1500 nm – 1610 nm. The output is split between two grating filters with adjustable bandwidth and center frequency followed by an erbium-doped fiber amplifier. The signal pulse centered at 1561 nm (group index $n_s = 10.3$ shown in figure 2(c)) with a duration of 10 ps and average power of 5 mW and the pump pulse centered at 1590 nm (group index $n_p = 4.2$) with 2 ps duration and 100 mW average power are launched into the waveguide successively, with a delay controlled by a line delay stage (General Photonics, MDL002). The pulse duration control is achieved by tuning the filter bandwidth and is checked by an autocorrelator (APE, pulseCheck 150). 0 ps delay is manually set when the maximal frequency shift of the signal peak position is observed. It should be mentioned that this zero definition does not mean that the signal and pump pulses enter the Bragg waveguide at the same time. The output light is collected and analyzed by an optical spectrum analyzer (Ando, AQ6317). The switch on/off of optical push broom effect is achieved by delaying pump pulse in respect to signal pulse, instead of turning on/off the pump pulse. In this way, the possible heat-induced perturbation of refractive index by pump illumination stays constant and could be ignored. More detailed information about the experimental setup can be found in support information.

## 3. Results and discussion

### 3.1. Frequency shift and spectral broadening

Figure 3 shows the signal spectra with the maximum frequency shift, which is obtained when the delay between signal and pump is vanishing, thus the signal travels inside the front for the maximum time and propagation distance. The black curve in Figure 3(a) represents the original spectrum serving as the baseline, which is obtained by applying a large delay to the pump pulse, such that the pump can never catch up with the signal before the signal leaves the waveguide at the rear grating coupler. Conversely, the red curve illustrates the spectrum's alteration attributable to the push broom effect, achieved by reducing the delay to an optimal value and setting it as 0 ps. Notably, the original signal's disappearance at 1561 nm is coupled with a frequency shift towards a lower wavelength and a broadening of the pulse spectrum. To mitigate the amplified spontaneous emission (ASE) noise emanating from both the pump and signal light, the data represented by the two curves in Figure 3(a) were transformed into a linear scale, subsequently subtracted, and the results are presented in Figure 3(b). This figure intuitively demonstrates the spectral shifting and broadening of the converted pulses within the 1550-1560 nm range. Furthermore, a minimal residual presence around 1561 nm suggests that nearly all signal energy have been effectively collected and converted to a lower wavelength.

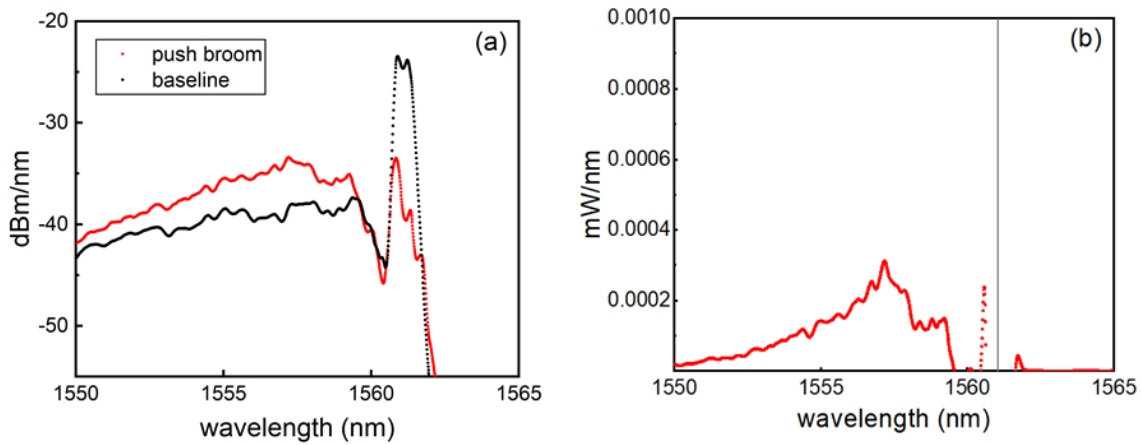

**Figure 3**. (a) Unaltered output signal at 1561 nm (black curve) while pumping at 1590 nm and applying a large optical delay such that the pump pulse does not catch up with the signal. Optical push broom takes place when shifting the pump pulse close behind the signal pulse at the waveguide input and allowing the pump to approach and collect the signal inside the waveguide (red curve). (b) To remove the ASE noise, the spectra in (a) were converted to linear scale, then the base line spectrum was subtracted from the pushbroomed signal spectrum. The vertical straight line denotes the original signal.

Furthermore, the dependency of the optical push broom effect on the delay between signal and front is explored by adjusting the timing of the pump pulse, as depicted in Figure 4(a). A 0 ps delay, defined as the point at which the maximum push broom effect occurs, causes the longest joint travel time and distance of signal and pump, thus yields the most pronounced frequency shift and spectral broadening. With increasing delay, the pump light enters the waveguide later, intersecting with the signal light only after the elapsed delay time. This results in a diminished duration for the two pulses to travel together, progressively weakening the frequency shift by push broom effect which is proportional to the time spent inside the slope of the refractive index perturbation [20]. The reduction in the frequency shift can be also explained by the Fourier transform induced by the optical push broom [12]. Three distinct delays—0 ps, 7 ps, and 20 ps—were chosen for analysis and are presented in Figure 4(a). These curves correspond to the signal pulse being fully converted, partially converted, and experiencing an almost negligible interaction, respectively. The dashed straight line represents the wavelength of the input pulse. An increase in delay leads to a reduction in frequency shift, underscoring the Fourier transform characteristics of the push broom effect, in alignment with theoretical predictions. Another indication of the partial conversion is observed in the residual power at the input wavelength, consisting of unconverted original signal components. This is more clearly illustrated by series measurements at different delays in support information Figure S4(a). The residual signal pulse at original wavelength is shorter than the original pulse and thus results into a slight broadening of the original signal spectrum which is observed as sharp peaks around the initial frequency in Figure 4(a).

In the case of optimal delay, the spectral broadening of the converted signal indicates the compression in time domain. However, a direct measurement of the compressed pulse is difficult due to its low power. An estimation of this compression will be given in the next section together with numerical simulations.

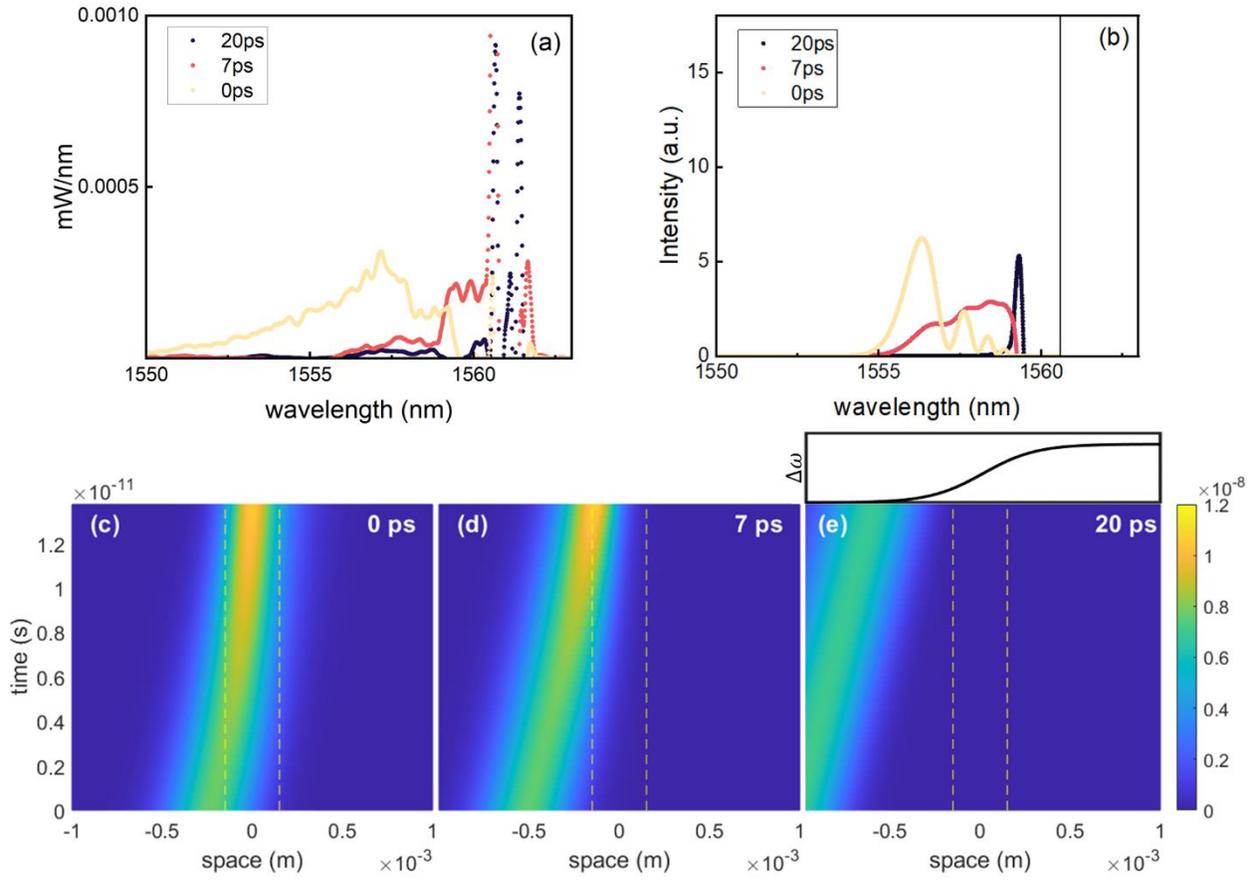

**Figure 4**. Signal wavelength distribution for different time delays between pump and signal from experiments (a) and corresponding SSFM calculations (b) prepared as subtraction of the converted and original signal. The delay of 0, 7, 20 ps represent a full conversion, partial conversion and nearly zero conversion, respectively. The oscillations in figure (b) for the delay of 0 ps origins from the signal light being partially located inside the front already at the input of the waveguide. The corresponding pulse evolution profile for three delays are shown in (c) – (e). The reference frame of the front is used, where the straight dashed line indicates the front width. In particular, the black curve on the top of figure (e) schematically shows the refractive index distribution of the front in space.

### 3.2. Numerical simulations of the optical push broom

To better interpret the experimental results, the numerical simulations for the pulse have been introduced and shown together in Fig 4(b) – (e) for comparison. The interactions between the signal light and moving index front was modelled using a slowly varying envelope approximation as follows [32–34]:

$$\frac{\partial A(t,z')}{\partial t} = (v_f - v_{g0})\frac{\partial A}{\partial z'} + \sum_{n=2}^{N} i^{(n-1)} \frac{\omega_n}{n!} \frac{\partial^n A}{\partial z'^n} + i\Delta\omega_D(z')A \qquad (1)$$

Here $A(t,z')$, $v_f$, $v_{g0}$ represent the signal envelope function, group velocity of the front and group velocity of the signal at wavenumber $\beta = 0$, respectively, where $\beta$ is defined in respect to the reference wavenumber. $\omega_n = \partial^n\omega/\partial\beta^n$ are the dispersion coefficients coming from the Taylor series expansion of the dispersion function $\omega(\beta)$ at $\beta = 0$ and $\Delta\omega_D(z')$ is the local shift of the dispersion relation due to the refractive index perturbation. The spatial coordinate $z' = z - v_f t$ represents the frame of the stationary front and $A(t, z')$ represents the temporal evolution of the signal envelope in space in respect to the front. The spectral distribution of the signal can be obtained by spatial Fourier transform of the envelope and recalculation from spatial frequency into temporal frequency [12].

This differential equation is solved numerically by split step Fourier method (SSFM) [35]. As mentioned above, the dispersion of the Bragg grating is approximated by the hyperbolic function $\omega(\beta) = \omega_{PBG} + \Delta\omega_{PBG} \cdot \sqrt{1 + [(\beta - \beta_{PBG})^2/\Delta\beta_{PBG}^2]}$, where $\omega_{PBG}$ and $\beta_{PBG}$ represent the central frequency and wavenumber of the bandgap, $\Delta\omega_{PBG}$ is the half of the bandgap width, and $\Delta\beta_{PBG} = \Delta\omega_{PBG}/v_{g\infty}$ is the parameter chosen to define the speed of light away from band gap as $v_{g\infty} = c/4.2$, where the hyperbolic function is converging to a straight line. The effect of this dispersion is calculated in reciprocal space as the phase accumulation with time according to $\omega(\beta)$ The hyperbolic dispersion curve is also confirmed by a numerical simulation of the band diagram in CST Studio Suite. The index front and the corresponding shift of the dispersion relation $\Delta\omega_D$ are described by a hyperbolic tangent function located at $z' = 0$: $\Delta\omega_D = \Delta\omega_{max}(1 + \tanh(z'/\Delta z_f))$, where $\Delta\omega_{max}$ and $\Delta z_f$ represents the pump induced maximum band diagram shift and front half width, receptively [36].

Three most important parameters for SSFM calculations are the maximal frequency shift of the dispersion relation due to pump excitation (0.4 THz), the group velocity of pump and signal. The detailed explanation for the parameters and related measurements are given in the support information. The conversion between temporal frequency distribution at position z, $B(\omega, z)$, and spatial frequency distribution at time $t$, $A(t, \beta)$, was achieved by considering the connection between linear energy density and power: $|B(\omega)|^2 = |A(\beta(\omega))|^2/v_g(\omega)$ [12]. Additionally, a phase shift $e^{i\,(\omega \cdot t_0 + \beta(\omega) \cdot z_0)}$ is multiplied to correct the phase to any chosen position $z_0$ for the given spatial envelope at time $t_0$. The reflection coefficient at input and output sections of the Bragg waveguide is approximated by group velocity mismatch $v_g(\omega)/v_f$ [37].

Building on the aforementioned analysis, the spectra at various delays have been computed and are presented in Figure 4(b). Notably, the oscillations—which are most pronounced at a 0 ps delay—stem from the interference caused by portions of the signal light initially positioned within the front at the waveguide entry. Similarly, 0 ps is defined as the delay at which the strongest frequency shift is observed. The observed slight shift in the signal wavelength, compared to experimental results, can be attributed to discrepancies between the estimated parameters and actual conditions. Despite these discrepancies, the calculations successfully replicate the experimental observations regarding frequency shift and spectral broadening. Further, pulse profile evaluations are depicted in Figures 4(c) to (e), offering an intuitive view of the push broom effect under different delay conditions. Here, the pulse is injected with 0, 7 and 20 ps delays and the dashed lines mark the front width. It becomes evident that an increase in delay leads to the signal light encountering the front later, resulting in diminished frequency shift and compression.

SSFM calculations provide a basis for estimating pulse compression attributable to the push broom effect. The compression factor is determined by the ratio of pulse durations at input and output ($\tau_{input}/\tau_{output}$). In this simulation, the maximum compression factor achieved is 2.5. This compression might not obvious from the Figure 4, as the signal should be first converted into the laboratory frame. Even if the signal pulse is not compressed in space, it is compressed in time as it is accelerated by the front.

Moreover, the maximal compression achievable in this waveguide can be estimated. Consistent with prior reports, the compression factor inversely correlates with the square of the initial signal duration [12]. The 10 ps experimental duration represents a balanced choice. While a longer duration would result in a stronger signal compression, it could obscure the delay dependence as large portion of the signal will be already inside the front at the input of the Bragg waveguide. Our choice of 10ps duration allows for a clear observation of the delay-dependent dynamics while still achieving adequate compression. However, for stronger compression, a longer permissible signal duration is preferred . The maximal possible compression duration for the signal in the front is defined as:

$$\tau_{\max} = L\left(\frac{|n_g^s - n_g^f|}{c}\right) \tag{2}$$

where $L$ is the length of the waveguide, $n_g^s$ and $n_g^f$ represents the group index of signal and front wavelengths. From this equation, a 20 ps signal pulse can be fully collected and compressed in a 1 mm waveguide. Consequently, the maximal compression factor under the inverse square law could be estimated to be 10 [12]. It's important to note that the above compression estimation is based on the spectral broadening from SSFM calculations only. Experimental data indicates a more significant broadening than predicted by SSFM calculations (see Figure 4(a) and (b), implying the above compression may be a conservative estimate. Notably, as addressed theoretically [12], in the case that one ideally applies an infinite waveguide with a non-decaying front, the signal would still be finitely compressed but continuously varying in wavevector. Consequently, the output duration would remain unchanged, while shifting to the further frequency.

## 3. Conclusion

In conclusion, we have successfully demonstrated the trapping and manipulation of a signal pulse within a moving index front inside a 1 mm long silicon Bragg waveguide. This innovative approach allowed for the compression of the signal pulse and its frequency shift to a higher spectrum, driven by a pump pulse-induced moving front. Additionally, the evolution of the pulse was accurately modeled using the slowly varying envelope approximation and numerically resolved through the split-step Fourier method. The direct measurement of the pulse durations after the push-broom setup was not possible due to the weak intensity of the transmitted signal pulse and mixing up with the pump pulse. We could well reproduce the experimentally observed frequency shifts with simulations and thus could estimate a compression of approx. 2.5 times. The potential for achieving even higher compression ratios exists and can be readily explored by employing a sharper moving front, extending the waveguide length and switching from lossy free carrier effect to Kerr nonlinearity. This research paves the way for advanced manipulation of light pulse temporal profiles, offering a streamlined, efficient pathway to one-step pulse compression.

**Acknowledgement**

We would like to acknowledge the sponsorship from Dassault Systems with their CST Studio Suite software. B. Z, M. A. G, M. E, and A. Yu. P would like to acknowledge the German Research Foundation (DFG) (project number: 392102174) for its financial support. H. L, X. C, and J. L acknowledge the support from National Natural Science Foundation of China (NSFC) (project number: 11761131001).

# Support Information

### Pulse Compression by an Optical Push Broom On a Chip

Boyi Zhang, Maurice Pfeiffer, Mahmoud A. Gaafar, He Li, Xinlun Cai, Juntao Li[*], Manfred Eich, and Alexander Yu. Petrov[†]

In this support information, the detailed experimental setup, SSFM simulation parameters and some additional experimental results are presented.

**Supplementary Note 1: Experimental Setup**

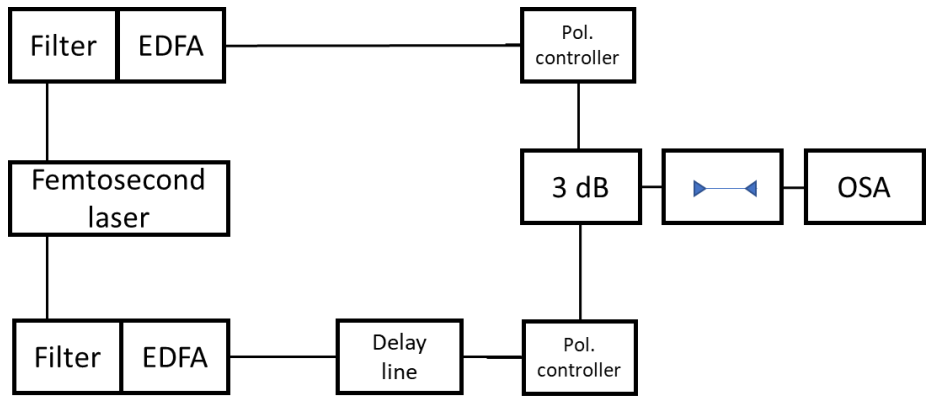

**Figure S1**: Schematic of the experimental setup. EDFA, erbium-doped fiber amplifier; Pol. controller, polarization controller and 3 dB, 50:50 beam combiner; OSA, optical spectrum analyzer. All solid lines represent fiber coupling.

The experiment is arranged as displayed in Figure S1. A mode-locked fiber laser (Menlo system) with a 100 MHz repetition rate delivers pulses of approx. 100 fs duration. The output is divided into pump and signal branches with adjustable bandwidth and center frequency (by filter, Santec, OTF-350) followed by an erbium-doped fiber amplifier (EDFA). 2-ps pump light centered at 1590 nm has an average power around 100 mW and polarization is controlled by a polarization controller (Thorlabs, FPC561) for an optimized coupling into the subsequent waveguide grating. Signal light is tunable from 1500 to 1590 nm with a duration ranging from 2 ps to 30 ps. Pulse duration is measured by an autocorrelator (APE, pulseCheck 150). The average power of the signal light after the EDFA is around 10 mW. Additionally, a delay line (GeneralPhotonics, MDL-002) is added after the EDFA to adjust the time delay of the signal pulse, which allows to control the delay within 550 ps. Subsequently, two pulses (pump and signal) are combined through a 50/50 (3 dB) beam combiner, which is then fiber-coupled to the grating coupler. The optical spectra are measured by the optical spectrum analyzer (Ando AQ6317). A SEM of the grating coupler is shown in Figure S2.

**Figure S2.** SEM image of the grating coupler.

The silicon Bragg waveguide and its coupler were fabricated on a Silicon-on-Insulator (SOI) substrate, featuring a 220 nm thick silicon layer. The patterning process began with the application of AR-P 6200.13 electron-beam photoresist, spin-coated onto the substrate. Subsequent patterning employed electron-beam lithography (EBL) to precisely define the waveguide structures. These patterns were then etched into the silicon layer using inductively coupled plasma (ICP) dry etching, ensuring accurate transfer of the intricate designs. Following this, residual photoresist was eliminated via oxygen plasma etching and a micro resist remover, preparing the surface for the final fabrication steps. The resulting grating couplers, designed for optimal efficiency in coupling between single-mode fibers and the silicon chip, feature a 20 µm length and 12 µm width. These couplers incorporate apodized photonic crystal columns to facilitate precise control over the coupling process [1,2]. Notably, the coupler achieves a 3dB bandwidth of 30 nm, centered at a peak wavelength of 1570 nm, while maintaining a minimal coupling loss of 7 dB per coupler.

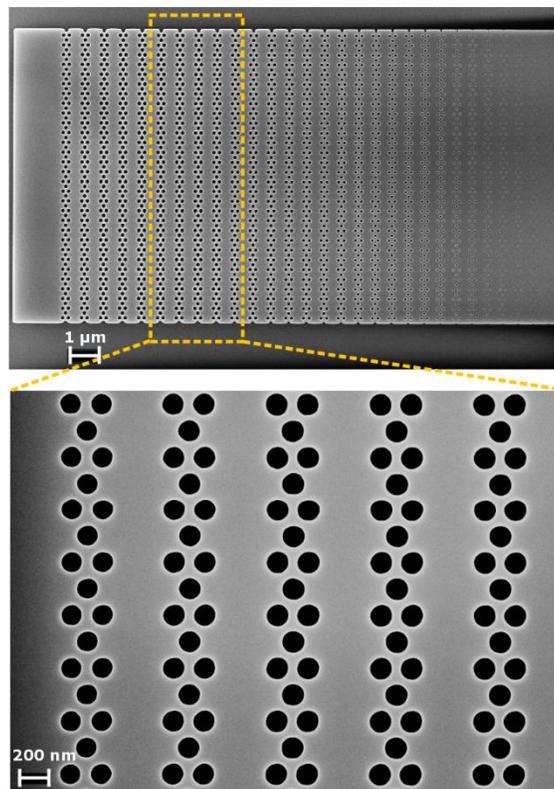

**Supplementary Note 2: Energy accumulation in the optical push broom effect**

To illustrate the energy accumulation characteristic of the optical push broom effect, we analyzed the spectrum resulting from cross-phase modulation (XPM), as depicted in Figure S3. In XPM experiments, the signal is shifted to 1547 nm, aligning the velocity of the signal pulse with that of the pump light. Consequently, the signal pulse 'surfs' on the index front rather than being collected and compressed. Figure S3 reveals that XPM induces frequency shifts to higher and lower frequencies, stemming from the signal pulse interacting with both the positive and negative slopes of the refractive index perturbation induced by the pump pulse.

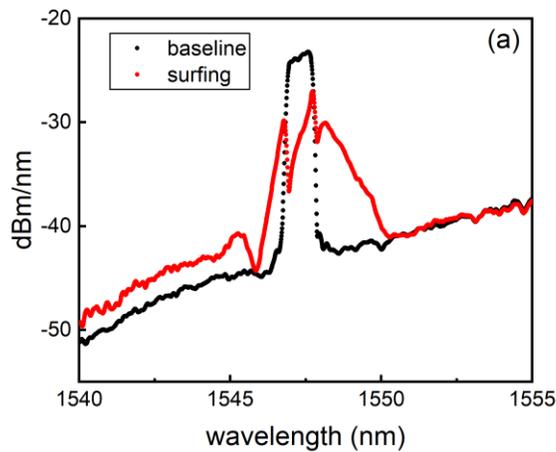

**Figure S3**. Signal spectra for the case when signal group velocity is similar to pump group velocity. The black line (baseline) is measured with a large delay between signal and pump so that they do not interact. The red line (surfing) represents the case when pump and signal co-propagate in the waveguide and the signal experiences XPM.

To elucidate the spectral changes more clearly, the spectra in Figure S3 were converted to a linear scale and subtracted, with the results presented in Figure S4(a). For comparative analysis, the push broom effect, discussed in the main text, is graphically represented alongside in Figure S4(b). The spectral efficiency on the y-axis—defined as the power per wavelength unit normalized to the transmitted power of the signal pulse at the output (1/nm)—facilitates a direct comparison between the conversion efficiencies of XPM and the push broom effect. Notably, the intensity scale for the push broom effect is expanded tenfold to accommodate its significantly higher intensity. The vertical line marking the input signal wavelength indicates that, in Figure S4(a), XPM generates a modest peak at 1546 nm. Conversely, the push broom effect produces a substantially broader peak with an intensity roughly 15 times greater, as shown in Figure S4(b). This marked increase in conversion efficiency underscores the ability of the push broom effect to accumulate energy, whereby the front traps and accelerates the signal pulse. Additionally, the residual signal at the input wavelength in the case of XPM, compared to the near-total conversion of the input signal to a new frequency under the push broom effect, further highlights the latter's efficiency.

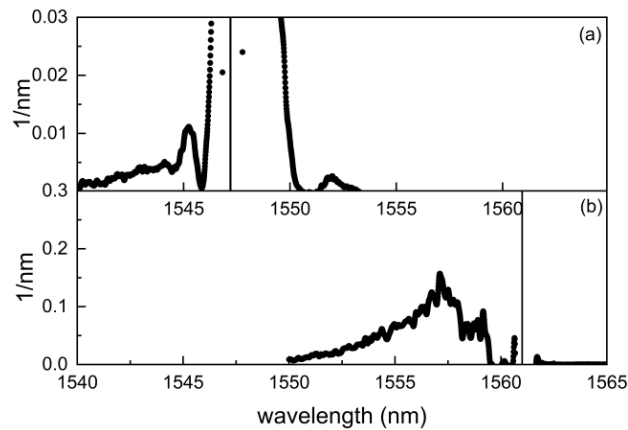

**Figure S4**. Comparison for the conversion efficiency between the XPM (a) and push broom effect (b). Note the intensity scale for the push broom effect (b) is expanded tenfold to accommodate its significantly higher intensity. The straight vertical line indicates the input signal wavelength.

**Supplementary Note 3: Optical push broom spectra vs delay**

In the main text, for the sake of simplicity, we presented three representative spectra corresponding to delays of 0 ps, 7 ps, and 20 ps. These were chosen to exemplify the signal pulse undergoing almost complete conversion, partial conversion, and negligible conversion. Additional spectra have been compiled and are displayed in Figure S5. Consistent with the presentation style of the main text, Figure S5(a) showcases the original spectra, while Figure S5(b) presents the results of linear subtraction. One can clearly see that the frequency shift by the push broom effect becomes weaker when increasing the delay between pump and signal pulses. Besides, the residual energy at the original signal wavelength is also increasing with growing delay.

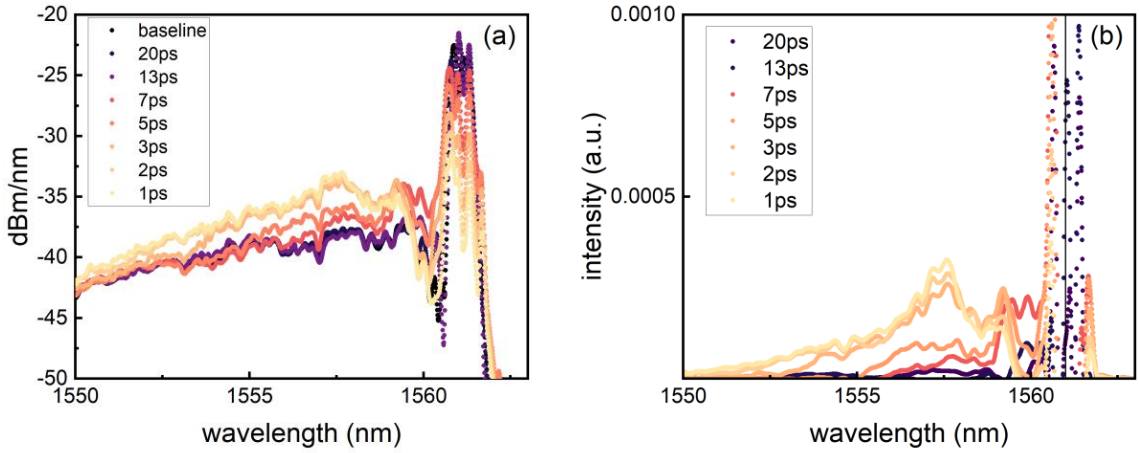

**Figure S5**. Signal spectra after interaction with the front at different time delay (a) and Difference between the signal spectra and the baseline in linear scale (b).

**Supplementary Note 4: Parameters for numerical simulations by SSFM**

The propagation of pulses along the waveguide is modelled by equation 1 and solved by split-step Fourier method numerically.

$$\frac{\partial A(t,z')}{\partial t} = (v_f - v_{g0})\frac{\partial A}{\partial z'} + \sum_{n=2}^{N} i^{(n-1)} \frac{\omega_n}{n!} \frac{\partial^n A}{\partial z'^n} + i\Delta\omega_D(z')A \tag{1}$$

Here $A(t,z)$, $v_{\text{f}}$, $v_{g0}$ represent the signal envelope function, group velocity of front, group velocity of signal at $\beta = 0$, respectively. In simulation, the signal is launched with a shifted $\beta$ value, by fitting the experimentally measured group velocity to the point on hyperbolic dispersion relation. The difference in speed of pump and signal light is measured indirectly by an optical oscilloscope. The front speed $v_{\text{f}}$ is obtained as a group velocity away from the bandgap simulated by the eigenmode solver of CST Studio Suite. The simulation gives the group index $n_{\text{f}}$ far away from the bandgap as 4.2. Then the central speed of the signal pulse at different wavelengths is evaluated by oscilloscope measurements. Thus, the simulation time (the time for pump light to pass the waveguide) is determined by pump speed and waveguide length.

Another critical parameter in our study is the frequency shift of the dispersion relation due to free carrier generation. This parameter essentially determines the extent of the frequency shift observed. While direct measurement poses challenges, we can approximate its magnitude by assessing the free carrier absorption effects. To this end, we analyzed the transmitted power of the signal light at the output of the waveguide across various delays [3], as depicted in Figure S5. Different from push broom effect, the signal light was set at 1546 nm where signal propagates with the same group velocity as pump and frequency changes are caused by cross phase modulation (XPM). It's important to note that the power contributions from the pump light have already been subtracted from these measurements. The strong drop of output power close to 0 delay corresponds to the two photon absorption (one photon from signal and one from pump) plus free carrier absorption, whereas the loss at larger negative delays indicates solely free carrier absorption. In such scenarios, the generated free carriers only increase the attenuation of the signal pulse and do not change its frequency. Based on the observed reduction in output power—approximately 25% in our experimental setup—we can infer the degree of free carrier absorption.

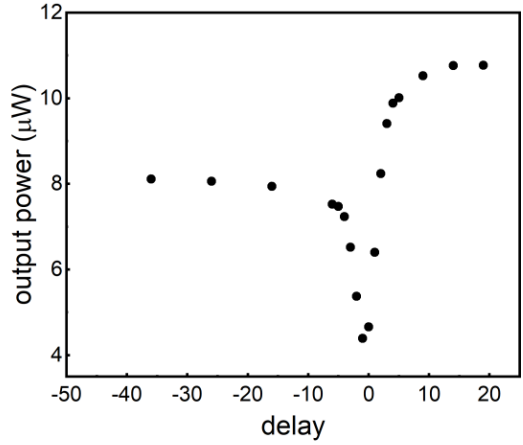

**Figure S5**. Transmitted signal power vs optical delay for signal at 1546 nm. 0 ps is set as a delay when maximal frequency shift by XPM is observed. Positive delay times represent the signal pulses that are running ahead of the pump, in which the drop to 0 delay corresponds to two photon absorption plus free carrier absorption. The transmission loss for negative delays compared with positive delays is solely caused by free carrier absorption, where pump is running ahead of signal.

The obtained absorption coefficient $\alpha$ can be used to estimate the generated free carrier concentration $\Delta N$:

$$\Delta \alpha = \sigma \Delta N \qquad (2)$$

where $\sigma = 1.45 \cdot 10^{-17}(\lambda/1.55)$ cm², $\lambda$ is the wavelength, $\Delta N$ represents the free carrier absorption cross-section, wavelength and free carrier concentration, respectively [3]. Afterwards, the change on refractive index could be expressed as [4]:

$$\Delta n = -(e^2 \lambda^2 / 8\pi^2 c^2 \varepsilon_0 n)[\Delta N_e / m^*_{ce} + \Delta N_h / m^*_{ch}] \qquad (3)$$

where e, $\varepsilon_0$, $n$ is the electron charge, permittivity of free space and refractive index of Si, respectively. $\Delta N_e$ and $m^*_{ce}$ represent the change of electron concentration and the effective mass of electrons, while $\Delta N_h$ and $m^*_{ch}$ correspond to the change of hole concentration and the effective mass of holes. $\Delta N_e$ and $\Delta N_h$ can be estimated from $\Delta N$ by $\Delta N_e = \Delta N^x$ and $\Delta N_h = \Delta N^y$. The coefficients *x* and *y* are experimentally fitted to be 1.04 and 0.818 respectively at 1550 nm in silicon by Soref et al [4]. They have also reported that at 1300 nm, $\Delta N_e = \Delta N^{1.05}$ and $\Delta N_h = \Delta N^{0.804}$. And in Dekker's calculation [3], $\Delta N_e = \Delta N^1$ and $\Delta N_h = \Delta N^{0.8}$ are applied at 1554 nm. It is important to emphasize that while the specific values chosen here do not alter the relative dependence of the push broom spectra on delay, they only makes the frequency shift stronger or weaker. This flexibility allows for the adjustment of the shifted peak's position or even for fitting simulation results to match experimental observations closely. After all, the maximal band diagram shift $\Delta\omega_{max}$ is estimated by:

$$\Delta\omega_{max}/\omega = \Delta n/n \qquad (4)$$

where $\Delta\omega_{max}$ stands for the frequency shifted by pump excitation. By adapting different electron/hole concentration coefficient *x* and *y*, $\Delta\omega_{max}$ varies within 0.161 – 0.486 THz. In our SSFM calculation, $\Delta\omega_{max} = 0.4$ THz is applied to roughly locate the spectra at experimental output.